\documentclass[
aps,
prd,
10pt,
twocolumn,
oneside,
preprintnumbers,
superscriptaddress,
amsmath,
amssymb,
nofootinbib,
flushbottom]{revtex4-2}

\usepackage{physics}
\usepackage{xspace}
\usepackage{slashed}
\usepackage{braket}
\usepackage{leftindex}
\usepackage{graphicx}  
\usepackage{bm}  
\usepackage{booktabs}
\usepackage[normalem]{ulem}
\usepackage{graphics}
\usepackage{makecell}
\usepackage{color}
\usepackage{colortbl}
\usepackage{hyperref}
\usepackage[svgnames,dvipsnames,table,x11names]{xcolor}
\usepackage{comment}
\usepackage{enumitem}
\usepackage{mathrsfs}

\hypersetup{
        pdfencoding=unicode,
	colorlinks=true,
	urlcolor=RoyalBlue,
	linkcolor=Maroon,
	citecolor=ForestGreen,
        pdftitle={Unitarity bounds and sum rules in the SMEFT},
	pdfauthor={},
	pdfdisplaydoctitle=true,
	pdfstartview=FitH
}

\makeatletter\g@addto@macro\bfseries{\boldmath}\makeatother

\begin{document}

\allowdisplaybreaks

\title{Unitarity bounds and sum rules in the SMEFT}

\author{Luigi~C.~Bresciani}
\email{luigicarlo.bresciani@phd.unipd.it}
\affiliation{Dipartimento di Fisica e Astronomia ``G.~Galilei'', Università degli Studi di Padova, Via F.~Marzolo 8,
I-35131 Padova, Italy}
\affiliation{Istituto Nazionale di Fisica Nucleare, Sezione di Padova, Via F.~Marzolo 8, I-35131 Padova, Italy}

\author{Paride~Paradisi}
\email{paride.paradisi@unipd.it}
\affiliation{Dipartimento di Fisica e Astronomia ``G.~Galilei'', Università degli Studi di Padova, Via F.~Marzolo 8,
I-35131 Padova, Italy}
\affiliation{Istituto Nazionale di Fisica Nucleare, Sezione di Padova, Via F.~Marzolo 8, I-35131 Padova, Italy}

\author{Andrea~Sainaghi}
\email{andrea.sainaghi@phd.unipd.it}
\affiliation{Dipartimento di Fisica e Astronomia ``G.~Galilei'', Università degli Studi di Padova, Via F.~Marzolo 8,
I-35131 Padova, Italy}
\affiliation{Istituto Nazionale di Fisica Nucleare, Sezione di Padova, Via F.~Marzolo 8, I-35131 Padova, Italy}
\affiliation{Physik-Institut, Universit\"at Z\"urich, Winterthurerstrasse 190, CH-8057 Z\"urich, Switzerland}

\begin{abstract}
We present a comprehensive reassessment of perturbative unitarity bounds in the dimension-six Standard Model Effective Field Theory, exploiting a new formalism based on spinor-helicity techniques to derive partial-wave unitarity bounds for generic $N \to M$ scattering amplitudes. We find that, in several cases, these theoretical constraints are already competitive with, or even stronger than, the corresponding experimental bounds for energy scales above a few TeV. This is especially the case for four-fermion operators under realistic flavor assumptions, where unitarity bounds can be further strengthened by exploiting sum rules. These results provide a valuable and timely contribution towards the ambitious program of new-physics searches with minimal theoretical priors supplemented by restrictive, yet general, first-principles criteria.

\end{abstract}

\maketitle

\section{Introduction}

The discovery of a scalar particle consistent with the Standard Model (SM) Higgs boson at the Large Hadron Collider (LHC) has firmly established the validity of the SM. Since no experimental data at present significantly contradict SM predictions, it is customary to assume the emergence of New Physics (NP) at energies much above the electroweak scale. In this context, Effective Field Theories (EFTs) provide a systematic framework to parameterize possible deviations from the SM by introducing higher-dimensional operators constructed from SM fields and respecting its symmetries (SMEFT) \cite{Buchmuller:1985jz,Grzadkowski:2010es,Brivio:2017vri,Isidori:2023pyp,Aebischer:2025qhh}. 

A key theoretical consistency requirement for any EFT is perturbative unitarity,
derived from the fundamental properties of the scattering S-matrix \cite{Jacob:1959at}. 
The resulting bounds are inherently perturbative, being derived at tree level, and
therefore do not represent sharp exclusions: their saturation signals the breakdown
of the low-energy description, indicating the energy scale at which new degrees of
freedom or a strongly coupled regime must appear.
Historically, unitarity considerations in $WW$ scattering led to an upper bound on the Higgs boson mass (the so-called no-lose Higgs theorem), which played a crucial role in motivating the construction of the LHC \cite{Lee:1977yc,Lee:1977eg}.
More recently, unitarity bounds have been extensively investigated in a wide variety of frameworks beyond the SM, including SMEFT, Higgs EFT, axion-like particle EFTs, and gravitational theories \cite{Gounaris:1994cm,Corbett:2014ora,Corbett:2017qgl,Cao:2023qks,Eboli:2026ccq,Degrande:2025uil,DiLuzio:2017chi,DiLuzio:2016sur,Mahmud:2024iyn,Allwicher:2021jkr,Cohen:2021ucp,Almeida:2020ylr,Cohen:2021gdw,Abu-Ajamieh:2020yqi,Brivio:2021fog,Falkowski:2019tft,Chang:2019vez,Alonso:2025ksv,Bresciani:2025ojh}. These studies provide constraints that are complementary to those obtained from direct and indirect experimental searches for NP particles.

Traditionally, unitarity bounds are derived from $2\to 2$ scattering processes using partial-wave decomposition and diagonalization of the scattering matrix \cite{Jacob:1959at}. While this approach has proven effective, it suffers from significant limitations: it cannot account for $2 \to N$ (with $N > 2$) processes, which are particularly relevant at the high energies accessible at the LHC and future colliders, nor is it easily applicable to spin-2 or higher-spin theories. 
By contrast, as recently demonstrated in \cite{Bresciani:2025toe}, on-shell methods---which have proved to be highly efficient in capturing the ultraviolet (UV) behavior of EFTs through renormalization group equations \cite{Caron-Huot:2016cwu,EliasMiro:2020tdv,Baratella:2020lzz,Jiang:2020mhe,Bern:2020ikv,Baratella:2020dvw,AccettulliHuber:2021uoa,EliasMiro:2021jgu,Baratella:2022nog,Machado:2022ozb,Bresciani:2023jsu,Bresciani:2024shu,Aebischer:2025zxg,Aebischer:2025ddl,Wu:2025qto}---also emerge as an ideal tool for deriving unitarity bounds for EFTs at high energies.

Partial-wave unitarity bounds within the SMEFT have already been partially explored in the literature.
In particular, Refs.~\cite{Corbett:2014ora,Corbett:2017qgl,Cao:2023qks} considered dimension-six bosonic operators and two-fermion-two-boson operators, but restricted the analysis to operators without gluon fields.
In these works, four-fermion operators were not considered.
Earlier studies addressed unitarity in specific four-fermion interactions \cite{Anders:1993dk}, but a systematic SMEFT treatment was still lacking.
Moreover, dimension-six operators $\psi^2\varphi^3$ modifying Yukawa interactions were neglected, and no unitarity bounds were derived for the Higgs operator $(\varphi^\dagger \varphi)^3$.
In this Letter, we fill these gaps by providing the first complete
coupled-channel implementation of perturbative unitarity for the
baryon-number-conserving dimension-six SMEFT. To this end, we compute
all relevant scattering amplitudes including every allowed helicity,
gauge, and flavor configuration, consistently retaining the correlations
among Wilson coefficients entering the same partial-wave amplitudes.
We then marginalize over each Wilson coefficient. 
This is especially relevant in flavor scenarios with dominant couplings to third-generation fermions, as naturally realized in models addressing the flavor puzzle at the TeV scale, which have attracted considerable interest in the recent literature~\cite{Bordone:2017bld,Greljo:2018tuh,Allwicher:2020esa,Fuentes-Martin:2020bnh,Fuentes-Martin:2022xnb,Barbieri:2023qpf,Barbieri:2024zkh,Davighi:2023evx,Davighi:2023iks,Fuentes-Martin:2024fpx,Covone:2024elw,Lizana:2024jby,FernandezNavarro:2023rhv,FernandezNavarro:2024hnv,FernandezNavarro:2025zmb,Isidori:2025rci,Greljo:2025mwj}.

The analyticity of the S-matrix allows one to derive spin-dependent sum rules \cite{Low:2009di,Altmannshofer:2023bfk,Altmannshofer:2025lun,Gu:2020thj,Azatov:2021ygj,Remmen:2020uze,Remmen:2022orj}, providing additional theoretical constraints on effective four-fermion interactions beyond partial-wave unitarity bounds. Remarkably, the resulting pattern of SMEFT inequalities can 
be used to infer characteristic features of the underlying NP dynamics. Therefore, a further goal of this Letter is to explore the complementarity of unitarity and sum-rule constraints, as well as their competitiveness with both current and projected experimental limits.

In the high-energy particle physics community, there is a growing consensus on performing NP searches with minimal theoretical priors. 
The huge and otherwise unmanageable parameter space must be supplemented by restrictive, yet general, first-principles criteria to make such searches tractable.
In this context, our work provides a systematic consistency framework that can be directly incorporated into present and future global SMEFT analyses.

\section{Theoretical bounds\label{sec:bounds}}

\paragraph{Partial-wave unitarity bounds.}
To compute the unitarity bounds, we adopt the method developed in \cite{Bresciani:2025toe}, which allows one to project a generic scattering amplitude $\ket{\mathcal{A}_{i\to f}}$ onto a kinematic basis $\ket{\mathcal{B}^J_{i\to f}}$ with definite angular momentum $J$, as follows:
\begin{equation}
\ket{\mathcal{A}_{i\to f}}=\sum_J a_{i\to f}^J \ket{\mathcal{B}_{i\to f}^J}\,,
\end{equation}
where $a_{i\to f}^J$ denote the partial-wave coefficients and the elements of the kinematic basis are normalized as 
\begin{equation}
    \langle \mathcal B^J_{i\to f}|\mathcal B^{J'}_{i\to f}\rangle = \int \dd\Phi_i \dd\Phi_f \, \mathcal B^{J'}_{i\to f} \left(\mathcal B^J_{i\to f}\right)^* = (2J+1)\delta^{JJ'}\,,
\end{equation}
where $\dd\Phi_{i,f}$ refers to the phase-space measure associated with the initial and final states.
In terms of $a_{i\to f}^J$, the partial-wave unitarity bounds 
take the form
\begin{equation}\label{unitarity bound general}
\abs{\Re a_{i\to i}^J}\leq 1\,,
\quad
0\leq \Im a_{i\to i}^J\leq 2\,,
\quad
\abs{a_{i\to f}^J}\leq 1\,,
\end{equation}
with $i \neq f$.
Further details on the construction of the basis and on the derivation of these bounds can be found in \cite{Bresciani:2025toe,Bresciani:2025ojh,Jiang:2020rwz,Shu:2021qlr}.

Using this framework, we systematically computed the unitarity bounds 
for all dimension-six SMEFT Wilson coefficients in the Warsaw basis \cite{Grzadkowski:2010es} reported in App.~\ref{app:ops}. 
The resulting bounds are summarized in Table \ref{tab:UniBounds}.

\begin{figure*}[h!t!b!]
\hspace{-3.mm}
    \includegraphics[width=0.26\textwidth]{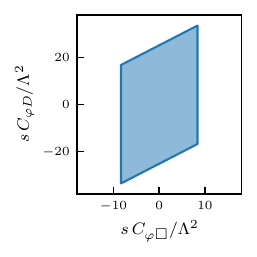}
    \hspace{-2.mm}
    \includegraphics[width=0.26\textwidth]{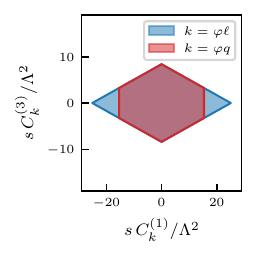}
    \hspace{-2.mm}
    \includegraphics[width=0.49\textwidth]{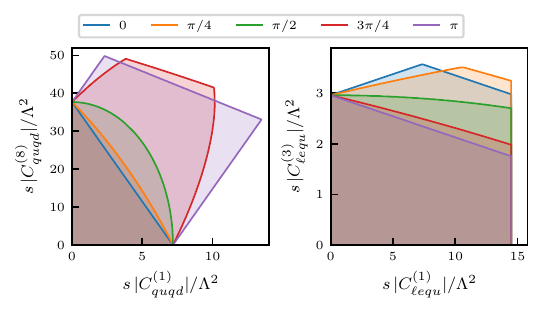}
    \caption{Parameter space allowed by unitarity bounds in the plane of Wilson coefficients in the single-flavor limit.
    In the last two plots, we also show the dependence on the relative complex phases $\mathrm{arg}(C_{quqd}^{(1)}/C_{quqd}^{(8)})$ and $\mathrm{arg}(C_{\ell equ}^{(1)}/C_{\ell equ}^{(3)})$.}
    \label{fig:uni3}
\end{figure*}

\begin{figure*}[h!t!b!]
    \hspace{-3.mm}
    \includegraphics[width=0.26\textwidth]{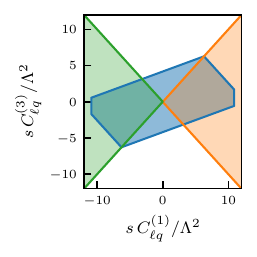}\hspace{-2.mm}%
    \includegraphics[width=0.26\textwidth]{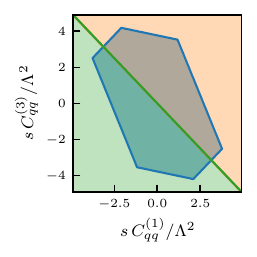}\hspace{-2.mm}%
    \includegraphics[width=0.26\textwidth]{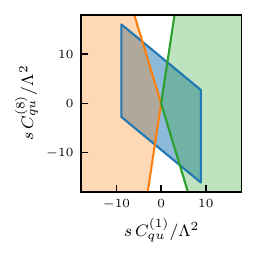}\hspace{-2.mm}%
    \includegraphics[width=0.26\textwidth]{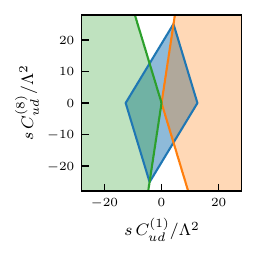}
    \caption{Parameter space allowed by unitarity bounds (blue) and spinning sum rules in the cases of scalar (orange) and vector (green) dominance, 
    in the single-flavor limit. The missing plot for the Wilson coefficients $C_{qd}^{(1,8)}$ is identical to that of $C_{qu}^{(1,8)}$.}
    \label{fig:1-flavor unitary+sum rules}
\end{figure*}

Several comments are in order. 
In contrast to the common practice in SMEFT analyses, where only one or a few operators are considered at a time, the unitarity bound associated with each Wilson coefficient is obtained by marginalizing over all other SMEFT coefficients. 
In practice, the marginalized bound on each coefficient is its largest value allowed by unitarity with all remaining coefficients free to vary. Since the partial-wave amplitudes depend linearly on the Wilson coefficients,
Eq.~\eqref{unitarity bound general} carves out a convex region in coefficient space,
and each marginalized bound is the projection of this region onto the corresponding
axis. Moreover, since all scattering channels are considered, the region is bounded. Indeed, by the amplitude/operator correspondence, the number of independent amplitudes equals that of independent operators, so that the unitarity conditions constrain every possible direction in coefficient space. Thus, the numbers in Table \ref{tab:UniBounds} should be interpreted as conservative one-dimensional projections of the full coupled-channel unitarity region, rather than as single-operator bounds.
This procedure is particularly relevant for coefficients that contribute to the same scattering amplitudes. This occurs for the following classes of operators: \textit{(i)} $(O_{\varphi \Box}, O_{\varphi D})$; \textit{(ii)} all pairs involving either a field-strength tensor $X$ or its dual $\widetilde X$; and \textit{(iii)} all pairs containing fermionic bilinears in the singlet or triplet (octet) representations of $\mathrm{SU}(2)$ ($\mathrm{SU}(3)$). Representative examples are shown in Figs.~\ref{fig:uni3} and \ref{fig:1-flavor unitary+sum rules}.

To fully exploit the power of this method, we performed a fully coupled-channel analysis for each operator. In particular, to compute the amplitude of a process mediated by a given SMEFT operator, we considered all possible initial and final states in all allowed helicity, gauge, and flavor configurations. 
At dimension six the leading numerical constraints often arise from the lowest-multiplicity channels. Nevertheless, the general $N \to M$ formalism \cite{Bresciani:2025toe} is essential for a uniform treatment of all operator classes, including $\varphi^6$ and $\psi^2\varphi^3$, and makes the inclusion of higher-multiplicity channels automatic
whenever they yield the leading energy growth.
For practical purposes, we present in Table \ref{tab:UniBounds} the unitarity bounds assuming single-flavor fermions, due to the large number of independent components arising from Wilson coefficients with multiple flavor indices.
In addition, the ancillary material includes an interactive Mathematica file containing both analytic and numerical unitarity bounds for four-fermion operators in the three-flavor scenario, under the minimal flavor violation and $\mathrm{U}(2)^5$ flavor assumptions~\cite{Faroughy:2020ina,Greljo:2022cah}.

Lastly, for operators of the type $X^3$ involving three gauge field strengths, the unitarity bounds were computed using amplitudes that include both the SM contribution and at most one SMEFT insertion, effectively truncating the calculation at $\mathcal{O}(1/\Lambda^2)$ at the amplitude level. 
Contributions of order $\mathcal{O}(1/\Lambda^4)$ could, in principle, yield stronger bounds, especially when the gauge coupling is small; however, including them without simultaneously accounting for dimension-eight operators would be inconsistent.
Indeed, from positivity constraints, the presence of a dimension-six operator of type $X^3$ necessarily implies the existence of a dimension-eight operator of type $X^4$, which would inevitably affect the resulting unitarity bounds; see \cite{Ghosh:2022qqq}.

Comparing the results in Table \ref{tab:UniBounds} with previous SMEFT analyses \cite{Corbett:2014ora,Corbett:2017qgl,Cao:2023qks}, we derive unitarity bounds for the complete set of four-fermion operators and for operators involving at least one gluon field, which were not covered in those works.
With this method, such bounds are straightforward to obtain compared to alternative approaches. Moreover, the fully exploited coupled-channel analysis allows us to derive stronger bounds than those previously reported. 

Overall, our bounds on purely bosonic operators agree with those in Ref.~\cite{Cao:2023qks}, except for a factor-of-two difference arising from Eq.~\eqref{unitarity bound general}, as that work conventionally imposed $|a_{i\to f}^J|\leq 2$ for the partial-wave coefficients. In contrast, the bounds on operators involving fermions exhibit larger discrepancies, which we attribute to the combined effects of the coupled-channel analysis and the marginalization procedure. In particular, operators of the type $\psi^2 X\varphi$ and $\psi^2\varphi^2 D$ show case-by-case differences of up to a factor of two. Overall, our bounds are always more stringent, except for $O_{\varphi ud}$, for which our bound is weaker by a factor of $\sqrt{2}$. Further details will be presented elsewhere.

\begin{table*}[h!t!b]
\renewcommand{\arraystretch}{1.5}
    \centering
    \begin{tabular}{c|c|c|c|c|c|c|c|c|c}
    \toprule
    \toprule
    \multicolumn{2}{c|}{{\boldmath$X^3$}} & \multicolumn{2}{c|}{{\boldmath$\varphi^6$} and {\boldmath$\varphi^4D^2$}} & \multicolumn{2}{c|}{{\boldmath$   \psi^2\varphi^3$}} & \multicolumn{2}{c|}{{\boldmath$(\overline{L}L)(\overline{L}L)$}} & \multicolumn{2}{c}{{\boldmath$(\overline{L}R)(\overline{R}L)$} and {\boldmath$(\overline{L}R)(\overline{L}R)$}}\\
    \hline
    $C_G$ & $4\pi/(9g_s)$ & $C_\varphi$ & $32\pi^3/3$ & $C_{e\varphi}$ & $32\pi^2/\sqrt{3}$ & $C_{\ell\ell}$ & $2\pi$ & $C_{\ell e dq}$ & $8\pi/\sqrt{3}$ \\
    
    $C_{\widetilde G}$ & $4\pi/(9g_s)$ & $C_{\varphi\Box}$ & $8\pi/3$ & $C_{u\varphi}$ & $32\pi^2/3$ & $C_{qq}^{(1)}$ & $6\pi/5$ & $C_{quqd}^{(1)}$ & $16\pi(1+\sqrt{2})/9$\\ 
        
    $C_W$ & $2\pi/(3g)$ &$C_{\varphi D}$ & $32\pi/3$ & $C_{d\varphi}$ & $32\pi^2/3$ & $C_{qq}^{(3)}$ & $4\pi/3$ & $C_{quqd}^{(8)}$ & $4\pi(2+7\sqrt{2})/3$\\
    
    $C_{\widetilde W}$ & $2\pi/(3g)$ & & & & & $C_{\ell q}^{(1)}$ & $2\pi\sqrt{3}$ & $C_{\ell equ}^{(1)}$ & $8\pi/\sqrt{3}$ \\

    &&&&&&  $C_{\ell q}^{(3)}$ & $2\pi$ & $C_{\ell e qu}^{(3)}$ & $(2+\sqrt{2})\pi/3$   \\
    
         \midrule
         \midrule
         \multicolumn{2}{c|}{{\boldmath$X^2\varphi^2$}} & \multicolumn{2}{c|}{{\boldmath$\psi^2 X \varphi$}} & \multicolumn{2}{c|}{{\boldmath$\psi^2\varphi^2D$}} & \multicolumn{2}{c|}{{\boldmath$(\overline{R}R)(\overline{R}R)$}} & \multicolumn{2}{c}{{\boldmath$(\overline{L}L)(\overline{R}R)$}}\\
         \hline
         $C_{\varphi G}$ & $\pi$ & $C_{eW}$ & $4\pi\sqrt{2/3}$ & $C_{\varphi \ell}^{(1)}$ & $8\pi$ & $C_{ee}$ & $2\pi$ & $C_{\ell e}$ & $4\pi$ \\
         
         $C_{\varphi \widetilde G}$ & $\pi$ & $C_{eB}$ & $4\pi$ & $C_{\varphi \ell}^{(3)}$ & $8\pi/3$ & $C_{uu}$ & $3\pi/2$ & $C_{\ell u}$ & $4\pi$\\
         
         $C_{\varphi W}$ & $2\pi\sqrt{2/3}$ & $C_{uG}$ & $2\pi\sqrt{3}$ & $C_{\varphi e}$ & $8\pi$ & $C_{dd}$ & $3\pi/2$ & $C_{\ell d}$ & $4\pi$\\
         
         $C_{\varphi \widetilde W}$ & $2\pi\sqrt{2/3}$ & $C_{uW}$ & $4\pi\sqrt{2/3}$ & $C_{\varphi q}^{(1)}$ & $2\pi\sqrt{6}$ & $C_{eu}$ & $4\pi$ & $C_{q e}$ & $4\pi$ \\
         
        $C_{\varphi B}$ & $2\pi\sqrt{2}$ & $C_{uB}$ & $4\pi$ & $C_{\varphi q}^{(3)}$ & $8\pi/3$ & $C_{ed}$ & $4\pi$ & $C_{q u}^{(1)}$ & $2\pi\sqrt{2}$\\
        
        $C_{\varphi \widetilde B}$ & $2\pi\sqrt{2}$ & $C_{dG}$ & $2\pi\sqrt{3}$ & $C_{\varphi u}$ & $4\pi \sqrt{3}$ & $C_{ud}^{(1)}$ & $4\pi$ & $C_{q u}^{(8)}$ & $3\pi(1+1/\sqrt{2})$\\
        
        $C_{\varphi W B}$ & $4\pi$ & $C_{dW}$ & $4\pi\sqrt{2/3}$ & $C_{\varphi d}$ & $4\pi\sqrt{3}$ & $C_{ud}^{(8)}$ & $8\pi$ & $C_{qd}^{(1)}$ & $2\pi\sqrt{2}$\\
        
        $C_{\varphi \widetilde W B}$ & $4\pi$ & $C_{dB}$ & $4\pi$ & $C_{\varphi ud}$ & $8\pi$ & & & $C_{q d}^{(8)}$ & $3\pi(1+1/\sqrt{2})$ \\
        \bottomrule
        \bottomrule
    \end{tabular}
    \caption{Unitarity bounds on the modulus of the baryon-number-conserving dimension-six SMEFT Wilson coefficients in the Warsaw basis \cite{Grzadkowski:2010es}, obtained by marginalizing over all other SMEFT coefficients (see Table \ref{tab:WarsawBasis}). The factors $s/\Lambda^2$ are left implicit, i.e., the reported values constrain the combination $s|C|/\Lambda^2$. Fermionic operators are considered in the single-flavor limit.}
    \label{tab:UniBounds}
\end{table*}

\paragraph{Spinning sum rules.}
The analyticity of the S-matrix, together with partial-wave unitarity, allows one to derive additional theoretical constraints on effective four-fermion interactions beyond the unitarity bounds discussed above. In particular, Refs.~\cite{Low:2009di,Azatov:2021ygj,Gu:2020thj,Remmen:2020uze,Remmen:2022orj,Altmannshofer:2023bfk,Altmannshofer:2025lun} have introduced new dispersion relations that yield either spin-dependent sum rules or constraints on the high-energy behavior of scattering amplitudes. Remarkably, the resulting pattern of SMEFT inequalities can be used to infer characteristic features of the underlying UV dynamics, indicating whether it is dominated by scalar or vector particles. A detailed derivation of such constraints is provided in Ref.~\cite{Remmen:2020uze}, where the authors used a basis that is different from the more conventional Warsaw basis. For this purpose, we report in App.~\ref{app: sum rules} some technical details on sum rules, as well as the corresponding inequalities regarding $\psi^4$ SMEFT Wilson coefficients in the Warsaw basis. 

As an illustration, Fig.~\ref{fig:1-flavor unitary+sum rules} shows the parameter space allowed by the unitarity bounds (blue) and the spinning sum rules under scalar (orange) and vector (green) dominance for four-fermion Wilson coefficients in the single-flavor limit. 
Stronger bounds generally arise in the three-flavor scenario and can be explored using the accompanying interactive Mathematica file.
We stress, however, that unlike the partial-wave unitarity bounds, which follow
solely from the consistency of the EFT, the spinning sum rules are considerably
more model-dependent and hold only under a specific set of assumptions whose
violation invalidates the corresponding inequalities
\cite{Remmen:2020uze,Remmen:2022orj,Azatov:2021ygj,Gu:2020thj}. In particular:
\textit{(i)} they assume tree-level UV completions and can be violated at one loop
\cite{Azatov:2021ygj}; \textit{(ii)} they assume dominance by states of
definite spin, so that completions involving a non-trivial interplay of spin-0 and
spin-1 states can evade them, since the relevant dispersion relation in
Eq.~\eqref{eq: sum rules master equation} ceases to be sign-definite; and
\textit{(iii)} they rely on the vanishing of boundary terms in the dispersion
relation, which fails in particular for vector $t$-channel UV completions
\cite{Remmen:2020uze,Remmen:2022orj,Azatov:2021ygj}.

This residual model dependence is, however, physically informative: a coefficient lying outside the scalar- and vector-dominated regions does not exclude a UV
completion but points to a more structured one, such as a vector $t$-channel mediator, 
a loop-level origin, or a mixed-spin spectrum. In App.~\ref{app:UV} we have verified this picture explicitly by matching to the dictionary of tree-level UV completions of Ref.~\cite{deBlas:2017xtg}: all scalar completions satisfy the sum-rule constraints, whereas vector $t$-channel completions can violate them, in agreement with limitation \textit{(iii)}.

\section{Phenomenology\label{sec:pheno}}

To assess the phenomenological impact of the theoretical bounds discussed above, it is useful to examine their complementarity with both current and anticipated future experimental constraints.

\begin{figure}[h!t!b!]
    \centering
    \includegraphics[]{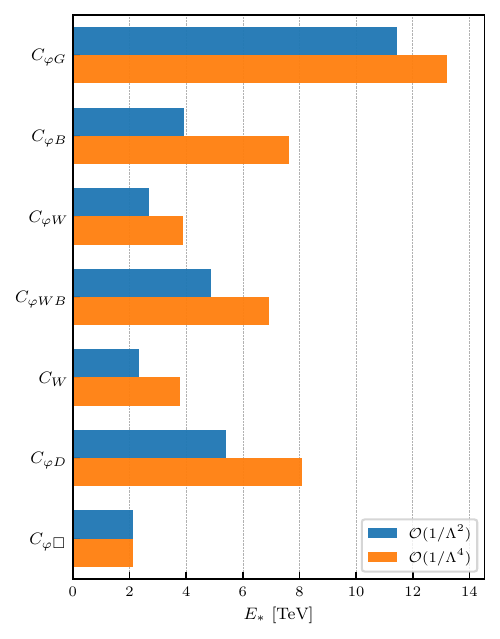}
    \caption{Minimum energy scales $E_*$ at which the unitarity bounds in Table \ref{tab:UniBounds} are violated for each purely bosonic Wilson coefficient in \cite{Celada:2024mcf}, using the 95\% CL marginalized intervals from linear (blue) and quadratic (orange) fits.}
    \label{fig:histogram}
\end{figure}

We first analyze the unitarity bounds for purely bosonic operators. Fig.~\ref{fig:histogram} shows the minimum energy scales $E_*$ at which the bounds in Table \ref{tab:UniBounds} are violated, assuming Wilson coefficients are taken from the 95\% CL marginalized intervals of Ref.~\cite{Celada:2024mcf}, using linear (blue) and quadratic (orange) fits.
\begin{figure*}[h!t!b!]
    \centering
    \includegraphics[width=0.329\textwidth]{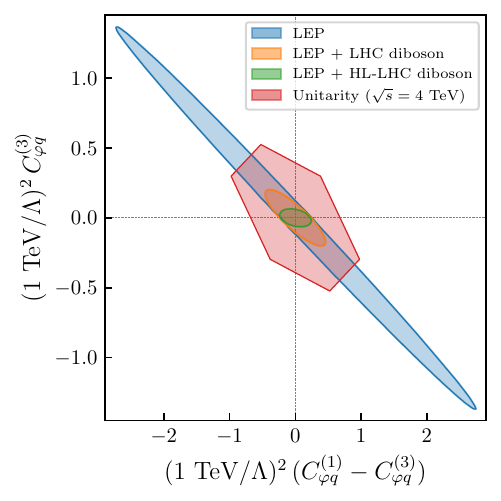}
    \hfill
    \includegraphics[width=0.329\textwidth]{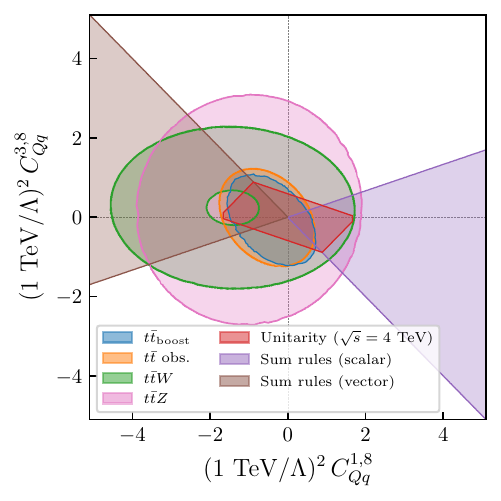}
    \hfill
    \includegraphics[width=0.329\textwidth]{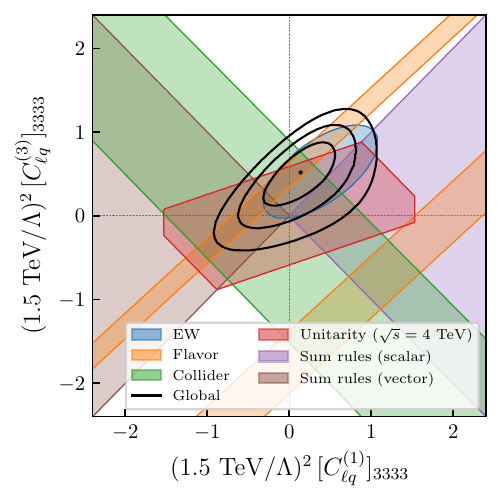}
    \caption{\textit{(Left)} Experimental and unitarity constraints on the Wilson coefficients of the operators $O_{\varphi q}^{(1)} = \sum_{p=1,2}i(\varphi^\dagger \overset{\leftrightarrow}{D}{}_\mu \varphi)(\overline q_p \gamma^\mu q_p)$ and $O_{\varphi q}^{(3)} = \sum_{p=1,2}i(\varphi^\dagger \overset{\leftrightarrow}{D}{}_\mu^I \varphi)(\overline q_p \sigma^I \gamma^\mu q_p)$. Experimental limits correspond to $95\%$ CL marginalized intervals \cite{Celada:2024mcf}.
    \textit{(Middle)} Experimental and unitarity constraints on the Wilson coefficients of the 
    operators $O_{Qq}^{1,8} = \sum_{p=1,2}(\overline q_3 \gamma_\mu T^A q_3)(\overline q_p \gamma^\mu T^A q_p)$ and $O_{Qq}^{3,8} = \sum_{p=1,2}(\overline q_3 \gamma_\mu T^A \sigma^I q_3)(\overline q_p \gamma^\mu T^A \sigma^I q_p)$. 
    Experimental limits correspond to $68\%$ CL and are derived from a quadratic fit \cite{Brivio:2019ius}. Here, the sum rules select the region where $C_{Qq}^{1,8}-(1\pm 2)C_{Qq}^{3,8}$ are positive (negative) for scalar (vector) completions.
    \textit{(Right)} Experimental and unitarity constraints on the Wilson coefficients of the 
    operators $[O_{\ell q}^{(1)}]_{3333} = (\overline \ell_3 \gamma^\mu \ell_3)(\overline q_3 \gamma_\mu q_3)$ and $[O_{\ell q}^{(3)}]_{3333} = (\overline \ell_3 \gamma^\mu \sigma^I \ell_3)(\overline q_3 \gamma_\mu \sigma^I q_3)$.
    Experimental limits correspond to $68\%$ CL \cite{Allwicher:2023shc}. These experimental contours are not recomputed by imposing an additional unitarity cut; the unitarity bounds should instead be interpreted as a theoretical consistency overlay.}
    \label{fig:3plots}
\end{figure*}
Moreover, Fig.~\ref{fig:3plots} (left) presents both the unitarity bounds and the experimental constraints from LEP and (HL-)LHC diboson data on the Wilson coefficients of the operators $O_{\varphi q}^{(1)}$ and $O_{\varphi q}^{(3)}$.
The experimental limits correspond to 95\% CL marginalized intervals obtained from a linear fit \cite{Celada:2024mcf}. As shown, the unitarity bounds---scaling as $(\sqrt{s}/4\,\mathrm{TeV})^2$---are significantly stronger than those from LEP alone, but remain weaker than the constraints derived from the combined LEP and (HL-)LHC data.

As a second example, we consider the global analysis of LHC Run II data in the top sector induced by the four-quark operators $O_{Qq}^{1,8}$ and $O_{Qq}^{3,8}$, where the experimental limits are obtained from a quadratic fit \cite{Brivio:2019ius}.
Most of the measurements involve final states with a top-quark pair, including kinematic distributions, the charge asymmetry, and associated top-pair production with a weak boson. As shown in Fig.~\ref{fig:3plots} (middle), the unitarity bounds are quite competitive with the LHC experimental limits. Interestingly, the inclusion of sum rules---corresponding to an underlying scalar or vector tree-level mediator of the four-fermion interactions---selects complementary regions of the parameter space, thereby providing insight into the nature of the underlying NP dynamics.

As a last example, we consider the comprehensive analysis of electroweak, flavor, and collider constraints on semileptonic 
four-fermion operators $[O_{\ell q}^{(1)}]_{3333}$ and $[O_{\ell q}^{(3)}]_{3333}$ in the $\mathrm{U}(2)^5$ scenario of Ref.~\cite{Allwicher:2023shc}, where NP predominantly couples to the third generation. Also in this case, as shown in Fig.~\ref{fig:3plots} (right), the unitarity bounds and the sum-rule constraints are highly complementary and competitive with the experimental limits.
In particular, it is noteworthy that the $1\,\sigma$ region of the global fit lies outside the regions preferred by the sum rules. 
If the current experimental situation is confirmed and further strengthened, this would favor, for instance, an underlying $t$-channel UV completion.

As a final remark, the unitarity bounds constrain the
combination $\hat s|C|/\Lambda^2$, so that the
corresponding limits become increasingly stringent at
larger values of the partonic center-of-mass energy
$\sqrt{\hat s}$. Since perturbative unitarity determines
the range of validity of the EFT, the relevant scale
entering the bound is $\sqrt{\hat s}$, rather than the
collider center-of-mass energy itself. This is
particularly relevant for the HL-LHC: although it
operates at the same collider energy as the LHC, its
much larger integrated luminosity enables precision
measurements deeper into the high-energy tails of
differential distributions, thereby extending the range
of partonic scales over which the SMEFT description can
be meaningfully tested. Accordingly, while most collider
constraints discussed here probe partonic
center-of-mass energies of order
$\sqrt{\hat s}\sim2~{\rm TeV}$, we display the unitarity
bounds at $\sqrt{\hat s}=4~{\rm TeV}$ as a
representative benchmark for the multi-TeV regime
accessible in HL-LHC measurements. In realistic
analyses with a broad partonic energy spectrum, the
unitarity condition should instead be evaluated at the
value of $\sqrt{\hat s}$ relevant for each event or
kinematic bin, thereby defining the maximum hard scale
up to which the SMEFT description remains
perturbatively self-consistent.

\section{Conclusions}

Non-renormalizable interactions induce scattering amplitudes that grow
with energy, potentially leading to violations of perturbative
unitarity at high scales.
A consistent interpretation of experimental
data, particularly at the (High-Luminosity) LHC and future high-energy
colliders, therefore requires quantitative theoretical criteria to
assess the range of validity of the SMEFT description.

In this Letter, we have performed a comprehensive reassessment of perturbative unitarity bounds on all dimension-six SMEFT Wilson coefficients (see Table~\ref{tab:UniBounds}), significantly extending and sharpening the partial results previously reported in the literature.
In particular, the unitarity bound associated with each Wilson coefficient was obtained here by marginalizing over all other SMEFT coefficients (see Figs.~\ref{fig:uni3} and \ref{fig:1-flavor unitary+sum rules}).
We performed a fully coupled-channel analysis, including all allowed initial and final states across helicity, gauge, and flavor configurations, thereby deriving stronger bounds than previously reported. 
We derived unitarity bounds
for the complete set of four-fermion operators as well as for operators
involving gluon fields, which were not covered in previous SMEFT
studies \cite{Corbett:2014ora,Corbett:2017qgl,Cao:2023qks}.

As shown in Fig.~\ref{fig:3plots}, these theoretical constraints are already competitive with---and in some cases stronger than---the corresponding experimental bounds at energy scales above a few TeV. 
This is particularly relevant for four-fermion operators under realistic flavor assumptions motivated by the SM flavor puzzle, where the unitarity limits can be further strengthened by imposing sum rules.

More broadly, the framework developed here establishes the first
systematic, perturbativity-based criterion that naturally incorporates
operator correlations and can be directly implemented in present and
future global SMEFT analyses, collider phenomenology, and Monte Carlo
event generation.
The results derived in this work strengthen the connection between theoretical and experimental constraints in future searches for physics beyond the Standard Model.

\begin{acknowledgments}
We would like to thank Alejo N.~Rossia for useful discussions and Gabriele Levati for valuable comments on the manuscript.
This work received funding by the INFN Iniziativa Specifica APINE and by the European Union's Horizon 2020 research and innovation programme under the Marie Sklodowska-Curie grant agreements n. 860881 --- HIDDeN, n. 101086085 --- ASYMMETRY. This work was also partially supported by the Italian MUR Departments of Excellence grant 2023-2027 ``Quantum Frontiers'' and by the European Union --- Next Generation EU and by the Italian Ministry of University and Research (MUR) via the PRIN 2022 project n. 2022K4B58X --- AxionOrigins. 
\end{acknowledgments}

\appendix

\section{SMEFT dimension-six operators\label{app:ops}}

In Table \ref{tab:WarsawBasis} we present the baryon-number-conserving dimension-six operators of the SMEFT in the Warsaw basis \cite{Grzadkowski:2010es}.
\begin{table*}[h!t!b]
\renewcommand{\arraystretch}{1.5}
    \centering
    \resizebox{\textwidth}{!}{
    \begin{tabular}{c | c|c | c|c| c|c| c|c| c}
    \toprule
    \toprule
    \multicolumn{2}{c|}{{\boldmath$X^3$}} & \multicolumn{2}{c|}{{\boldmath$\varphi^6$} and {\boldmath$\varphi^4D^2$}} & \multicolumn{2}{c|}{{\boldmath$\psi^2\varphi^3$}} & \multicolumn{2}{c|}{{\boldmath$(\overline{L}L)(\overline{L}L)$}} & \multicolumn{2}{c}{{\boldmath$(\overline{L}R)(\overline{R}L)$} and {\boldmath$(\overline{L}R)(\overline{L}R)$}} \\
    \hline
    $O_G$ & $f^{ABC} G_\mu^{A\nu} G_\nu^{B\rho} G_\rho^{C\mu}$ & $O_\varphi$ & $(\varphi^\dagger \varphi)^3$ & $^\ddagger O_{e\varphi}$ & $(\varphi^\dagger \varphi)(\overline{\ell}_p e_r \varphi)$ & $O_{\ell\ell}$ & $(\overline{\ell}_p \gamma_\mu \ell_r)(\overline{\ell}_s \gamma^\mu \ell_t)$ & $^\ddagger O_{\ell edq}$ & $(\overline{\ell}_p^j e_r)(\overline{d}_s q_{tj})$ \\
    
    $O_{\widetilde{G}}$ & $f^{ABC} \widetilde{G}_\mu^{A\nu} G_\nu^{B\rho} G_\rho^{C\mu}$ & $O_{\varphi\Box}$ & $(\varphi^\dagger \varphi)\Box(\varphi^\dagger \varphi)$ & $^\ddagger O_{u\varphi}$ & $(\varphi^\dagger \varphi)(\overline{q}_p u_r \widetilde{\varphi})$ & $O_{qq}^{(1)}$ & $(\overline{q}_p \gamma_\mu q_r)(\overline{q}_s \gamma^\mu q_t)$ & $^\ddagger O_{quqd}^{(1)}$ & $(\overline{q}_p^j u_r)\epsilon_{jk}(\overline{q}_s^k d_t)$ \\
    
    $O_W$ & $\epsilon^{IJK} W_\mu^{I\nu} W_\nu^{J\rho} W_\rho^{K\mu}$ & $O_{\varphi D}$ & $(\varphi^\dagger D^\mu \varphi)^* (\varphi^\dagger D_\mu \varphi)$ & $^\ddagger O_{d\varphi}$ & $(\varphi^\dagger \varphi)(\overline{q}_p d_r \varphi)$ & $O_{qq}^{(3)}$ & $(\overline{q}_p \gamma_\mu \sigma^I q_r)(\overline{q}_s \gamma^\mu \sigma^I q_t)$ & $^\ddagger O_{quqd}^{(8)}$ & $(\overline{q}_p^j T^A u_r)\epsilon_{jk}(\overline{q}_s^k T^A d_t)$ \\
    
    $O_{\widetilde{W}}$ & $\epsilon^{IJK} \widetilde{W}_\mu^{I\nu} W_\nu^{J\rho} W_\rho^{K\mu}$ & & & & & $O_{\ell q}^{(1)}$ & $(\overline{\ell}_p \gamma_\mu \ell_r)(\overline{q}_s \gamma^\mu q_t)$ & $^\ddagger O_{\ell equ}^{(1)}$ & $(\overline{\ell}_p^j e_r)\epsilon_{jk}(\overline{q}_s^k u_t)$ \\
    
    & & & & & & $O_{\ell q}^{(3)}$ & $(\overline{\ell}_p \gamma_\mu \sigma^I \ell_r)(\overline{q}_s \gamma^\mu \sigma^I q_t)$ & $^\ddagger O_{\ell equ}^{(3)}$ & $(\overline{\ell}_p^j \sigma_{\mu\nu} e_r)\epsilon_{jk}(\overline{q}_s^k \sigma^{\mu\nu} u_t)$ \\
    
    \midrule
    \midrule
    \multicolumn{2}{c|}{{\boldmath$X^2\varphi^2$}} & \multicolumn{2}{c|}{{\boldmath$\psi^2 X \varphi$}} & \multicolumn{2}{c|}{{\boldmath$\psi^2\varphi^2D$}} & \multicolumn{2}{c|}{{\boldmath$(\overline{R}R)(\overline{R}R)$}} & \multicolumn{2}{c}{{\boldmath$(\overline{L}L)(\overline{R}R)$}} \\
    \hline
    $O_{\varphi G}$ & $(\varphi^\dagger \varphi) G_{\mu\nu}^A G^{A\mu\nu}$ & $^\ddagger O_{eW}$ & $(\overline{\ell}_p \sigma^{\mu\nu} e_r)\sigma^I \varphi W_{\mu\nu}^I$ & $O_{\varphi\ell}^{(1)}$ & $(\varphi^\dagger i\overset{\leftrightarrow}{D}{}_\mu \varphi)(\overline{\ell}_p \gamma^\mu \ell_r)$ & $O_{ee}$ & $(\overline{e}_p \gamma_\mu e_r)(\overline{e}_s \gamma^\mu e_t)$ & $O_{\ell e}$ & $(\overline{\ell}_p \gamma_\mu \ell_r)(\overline{e}_s \gamma^\mu e_t)$ \\
    
    $O_{\varphi \widetilde{G}}$ & $(\varphi^\dagger \varphi) \widetilde{G}_{\mu\nu}^A G^{A\mu\nu}$ & $^\ddagger O_{eB}$ & $(\overline{\ell}_p \sigma^{\mu\nu} e_r)\varphi B_{\mu\nu}$ & $O_{\varphi\ell}^{(3)}$ & $(\varphi^\dagger i\overset{\leftrightarrow}{D}{}_\mu^I \varphi)(\overline{\ell}_p \sigma^I \gamma^\mu \ell_r)$ & $O_{uu}$ & $(\overline{u}_p \gamma_\mu u_r)(\overline{u}_s \gamma^\mu u_t)$ & $O_{\ell u}$ & $(\overline{\ell}_p \gamma_\mu \ell_r)(\overline{u}_s \gamma^\mu u_t)$ \\
    
    $O_{\varphi W}$ & $(\varphi^\dagger \varphi) W_{\mu\nu}^I W^{I\mu\nu}$ & $^\ddagger O_{uG}$ & $(\overline{q}_p \sigma^{\mu\nu} T^A u_r)\widetilde{\varphi} G_{\mu\nu}^A$ & $O_{\varphi e}$ & $(\varphi^\dagger i\overset{\leftrightarrow}{D}{}_\mu \varphi)(\overline{e}_p \gamma^\mu e_r)$ & $O_{dd}$ & $(\overline{d}_p \gamma_\mu d_r)(\overline{d}_s \gamma^\mu d_t)$ & $O_{\ell d}$ & $(\overline{\ell}_p \gamma_\mu \ell_r)(\overline{d}_s \gamma^\mu d_t)$ \\
    
    $O_{\varphi \widetilde{W}}$ & $(\varphi^\dagger \varphi) \widetilde{W}_{\mu\nu}^I W^{I\mu\nu}$ & $^\ddagger O_{uW}$ & $(\overline{q}_p \sigma^{\mu\nu} u_r)\sigma^I \widetilde{\varphi} W_{\mu\nu}^I$ & $O_{\varphi q}^{(1)}$ & $(\varphi^\dagger i\overset{\leftrightarrow}{D}{}_\mu \varphi)(\overline{q}_p \gamma^\mu q_r)$ & $O_{eu}$ & $(\overline{e}_p \gamma_\mu e_r)(\overline{u}_s \gamma^\mu u_t)$ & $O_{qe}$ & $(\overline{q}_p \gamma_\mu q_r)(\overline{e}_s \gamma^\mu e_t)$ \\
    
    $O_{\varphi B}$ & $(\varphi^\dagger \varphi) B_{\mu\nu} B^{\mu\nu}$ & $^\ddagger O_{uB}$ & $(\overline{q}_p \sigma^{\mu\nu} u_r)\widetilde{\varphi} B_{\mu\nu}$ & $O_{\varphi q}^{(3)}$ & $(\varphi^\dagger i\overset{\leftrightarrow}{D}{}_\mu^I \varphi)(\overline{q}_p \sigma^I \gamma^\mu q_r)$ & $O_{ed}$ & $(\overline{e}_p \gamma_\mu e_r)(\overline{d}_s \gamma^\mu d_t)$ & $O_{qu}^{(1)}$ & $(\overline{q}_p \gamma_\mu q_r)(\overline{u}_s \gamma^\mu u_t)$ \\
    
    $O_{\varphi \widetilde{B}}$ & $(\varphi^\dagger \varphi) \widetilde{B}_{\mu\nu} B^{\mu\nu}$ & $^\ddagger O_{dG}$ & $(\overline{q}_p \sigma^{\mu\nu} T^A d_r)\varphi G_{\mu\nu}^A$ & $O_{\varphi u}$ & $(\varphi^\dagger i\overset{\leftrightarrow}{D}{}_\mu \varphi)(\overline{u}_p \gamma^\mu u_r)$ & $O_{ud}^{(1)}$ & $(\overline{u}_p \gamma_\mu u_r)(\overline{d}_s \gamma^\mu d_t)$ & $O_{qu}^{(8)}$ & $(\overline{q}_p \gamma_\mu T^A q_r)(\overline{u}_s \gamma^\mu T^A u_t)$ \\
    
    $O_{\varphi WB}$ & $(\varphi^\dagger \sigma^I \varphi) W_{\mu\nu}^I B^{\mu\nu}$ & $^\ddagger O_{dW}$ & $(\overline{q}_p \sigma^{\mu\nu} d_r)\sigma^I \varphi W_{\mu\nu}^I$ & $O_{\varphi d}$ & $(\varphi^\dagger i\overset{\leftrightarrow}{D}{}_\mu \varphi)(\overline{d}_p \gamma^\mu d_r)$ & $O_{ud}^{(8)}$ & $(\overline{u}_p \gamma_\mu T^A u_r)(\overline{d}_s \gamma^\mu T^A d_t)$ & $O_{qd}^{(1)}$ & $(\overline{q}_p \gamma_\mu q_r)(\overline{d}_s \gamma^\mu d_t)$ \\
    
    $O_{\varphi \widetilde{W}B}$ & $(\varphi^\dagger \sigma^I \varphi) \widetilde{W}_{\mu\nu}^I B^{\mu\nu}$ & $^\ddagger O_{dB}$ & $(\overline{q}_p \sigma^{\mu\nu} d_r)\varphi B_{\mu\nu}$ & $^\ddagger O_{\varphi ud}$ & $(\widetilde{\varphi}^\dagger i D_\mu \varphi)(\overline{u}_p \gamma^\mu d_r)$ & & & $O_{qd}^{(8)}$ & $(\overline{q}_p \gamma_\mu T^A q_r)(\overline{d}_s \gamma^\mu T^A d_t)$ \\
    \bottomrule
    \bottomrule
    \end{tabular}
    }
    \caption{Dimension-six operators of the SMEFT in the Warsaw basis \cite{Grzadkowski:2010es} that conserve baryon number. Non-Hermitian operators are denoted with $^\ddagger O$ and the corresponding Hermitian conjugates must be included in the Lagrangian.}
    \label{tab:WarsawBasis}
\end{table*}
The relevant conventions are as follows: $\widetilde \varphi^j = \epsilon_{jk} (\varphi^k)^*$ (with $\epsilon_{12}=1$), $\varphi^\dagger \overset{\leftrightarrow}{D}{}_\mu \varphi = \varphi^\dagger D_\mu \varphi - (D_\mu \varphi)^\dagger \varphi$ and $\varphi^\dagger \overset{\leftrightarrow}{D}{}_\mu^I \varphi = \varphi^\dagger \sigma^I D_\mu \varphi - (D_\mu \varphi)^\dagger \sigma^I \varphi$, dual field-strength tensors are defined as $\widetilde X_{\mu\nu} = \frac{1}{2}\epsilon_{\mu\nu\rho\sigma}X^{\rho\sigma}$ (with $X=B,W^I,G^A$ and $\epsilon_{0123} = 1$), $\lambda^A_{\alpha\beta}$ are the Gell-Mann matrices and $\sigma^I_{jk}$ are the Pauli matrices, $T_{\alpha\beta}^A = \frac{1}{2}\lambda^A_{\alpha\beta}$ are the $\mathrm{SU}(3)$ generators, and $\sigma_{\mu\nu} = \frac{i}{2}[\gamma_\mu,\gamma_\nu]$.
The metric, whose signature is relevant for the spinning sum rules discussed in App.~\ref{app: sum rules}, is $g_{\mu\nu} = \mathrm{diag}(+1,-1,-1,-1)$.

\section{Further details on spinning sum rules\label{app: sum rules}}

\begin{table}[h!t!b!]
\renewcommand{\arraystretch}{1.5}
    \centering
    \resizebox{\columnwidth}{!}{
    \begin{tabular}{cc}
    \toprule
    \toprule
    {Operator class} & {Spinning sum rules} \\
    \midrule
    {\boldmath$(\overline{R}R)(\overline{R}R)$} &
    $[C_{ee}]_{prst}$ \\
    & $[C_{uu}]_{prst}$\,, \qquad $[C_{uu}]_{prst} + [C_{uu}]_{ptsr}$ \\
    & $[C_{dd}]_{prst}$\,, \qquad $[C_{dd}]_{prst} + [C_{dd}]_{ptsr}$ \\
    & $[C_{eu}]_{prst}$\,, \qquad $[C_{ed}]_{prst}$ \\
    & $[C_{ud}^{(1)}]_{prst}+\tfrac{1}{3}[C_{ud}^{(8)}]_{prst}$ \\
    & $[C_{ud}^{(1)}]_{prst}-\tfrac{1}{6}[C_{ud}^{(8)}]_{prst}$ \\
    \midrule
    {\boldmath$(\overline{L}L)(\overline{L}L)$} &
    $[C_{\ell\ell}]_{prst}$\,, \qquad $[C_{\ell\ell}]_{prst} + [C_{\ell\ell}]_{ptsr}$ \\
    & $[C_{qq}^{(1)}]_{prst} + [C_{qq}^{(1)}]_{ptsr} + [C_{qq}^{(3)}]_{prst} + [C_{qq}^{(3)}]_{ptsr}$ \\
    & $[C_{qq}^{(1)}]_{prst} - [C_{qq}^{(3)}]_{prst} + 2[C_{qq}^{(3)}]_{ptsr}$ \\
    & $[C_{\ell q}^{(1)}]_{prst} + [C_{\ell q}^{(3)}]_{prst}$ \\
    & $[C_{\ell q}^{(1)}]_{prst} - [C_{\ell q}^{(3)}]_{prst}$ \\
    \midrule
    {\boldmath$(\overline{L}L)(\overline{R}R)$} &
    $-[C_{\ell e}]_{prst}$\,, \qquad $-[C_{\ell u}]_{prst}$ \\
    & $-[C_{\ell d}]_{prst}$\,, \qquad $-[C_{qe}]_{prst}$ \\
    & $-[C_{qu}^{(1)}]_{prst} - \tfrac{1}{3}[C_{qu}^{(8)}]_{prst}$ \\
    & $-[C_{qu}^{(1)}]_{prst} + \tfrac{1}{6}[C_{qu}^{(8)}]_{prst}$ \\
    & $-[C_{qd}^{(1)}]_{prst} - \tfrac{1}{3}[C_{qd}^{(8)}]_{prst}$ \\
    & $-[C_{qd}^{(1)}]_{prst} + \tfrac{1}{6}[C_{qd}^{(8)}]_{prst}$ \\
    \bottomrule
    \bottomrule
    \end{tabular}
    }
    \caption{Combinations of four-fermion Wilson coefficients that, when contracted with $\alpha_p \alpha_r^* \beta^*_s \beta_t$, are positive (negative) if the UV completion is dominated by scalars (vectors).}
    \label{tab:sum_rules}
\end{table}

\begin{table*}[h!t!b!]
\renewcommand{\arraystretch}{1}
    \centering
    \begin{tabular}{cccccccc}
    \toprule
    \toprule
    Field & Irrep & Spin & Channel & $[C_{\ell q}^{(1)}]_{3333}/\Lambda^2$ & $[C_{\ell q}^{(3)}]_{3333}/\Lambda^2$ & Unitarity bound & Plot \\
    \midrule
    $\omega_1$ & $(3,1)_{-\frac{1}{3}}$ & $0$ & $s$ & $|[y^{q\ell}_{\omega_1}]_{33}|^2/(4M^2_{\omega_1})$ & $-|[y^{q\ell}_{\omega_1}]_{33}|^2/(4M^2_{\omega_1})$ & $|[y^{q\ell}_{\omega_1}]_{33}|^2\,s\leq 4\pi M^2_{\omega_1}$ & \makecell[c]{\includegraphics[scale = 0.6]{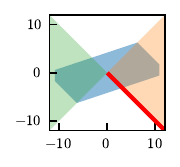}} \\
    $\zeta$ & $(3,3)_{-\frac{1}{3}}$ & $0$ & $s$ & $3|[y^{q\ell}_{\zeta}]_{33}|^2/(4M^2_{\zeta})$ & $|[y^{q\ell}_{\zeta}]_{33}|^2/(4M^2_{\zeta})$ & $|[y^{q\ell}_{\zeta}]_{33}|^2\,s\leq 4\pi M^2_{\zeta}$ & \makecell[c]{\includegraphics[scale = 0.6]{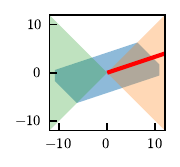}} \\
    $\mathcal{U}_2$ & $(3,1)_{\frac{2}{3}}$ & $1$ & $s$ & $-|[g_{\mathcal U_2}^{\ell q}]_{33}|^2/(2M^2_{\mathcal U_2})$ & $-|[g_{\mathcal U_2}^{\ell q}]_{33}|^2/(2M^2_{\mathcal U_2})$ & $|[g_{\mathcal U_2}^{\ell q}]_{33}|^2\,s\leq 4\pi M^2_{\mathcal U_2}$ & \makecell[c]{\includegraphics[scale = 0.6]{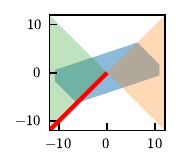}} \\
    $\mathcal{X}$ & $(3,3)_{\frac{2}{3}}$ & $1$ & $s$ & $-3|[g_{\mathcal X}]_{33}|^2/(8M^2_{\mathcal X})$ & $|[g_{\mathcal X}]_{33}|^2/(8M^2_{\mathcal X})$ & $|[g_{\mathcal X}]_{33}|^2\,s\leq \tfrac{16}{3}\pi M^2_{\mathcal X}$ & \makecell[c]{\includegraphics[scale = 0.6]{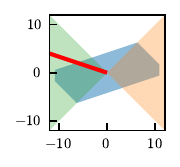}} \\
    $\mathcal{B}$ & $(1,1)_{0}$ & $1$ & $t$ & $-[g_{\mathcal B}^\ell]_{33}[g_{\mathcal B}^q]_{33}/M^2_{\mathcal B}$ & $0$ & $|[g_{\mathcal B}^\ell]_{33}[g_{\mathcal B}^q]_{33}|\,s\leq 2\sqrt{3}\,\pi M^2_{\mathcal B}$ & \makecell[c]{\includegraphics[scale = 0.6]{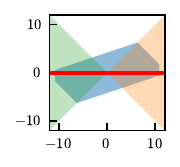}} \\
    $\mathcal{W}$ & $(1,3)_{0}$ & $1$ & $t$ & $0$ & $-[g_{\mathcal W}^\ell]_{33}[g_{\mathcal W}^q]_{33}/(4M^2_{\mathcal W})$ & $|[g_{\mathcal W}^\ell]_{33}[g_{\mathcal W}^q]_{33}|\,s\leq \tfrac{16}{3}\pi M^2_{\mathcal W}$ & \makecell[c]{\includegraphics[scale = 0.6]{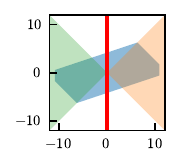}} \\
    \bottomrule
    \bottomrule
    \end{tabular}
    \caption{Tree-level matching of single-field UV completions \cite{deBlas:2017xtg}
    onto $[O_{\ell q}^{(1)}]_{3333}$ and $[O_{\ell q}^{(3)}]_{3333}$. The penultimate
    column reports the associated perturbative unitarity bound, obtained as the most
    stringent of the constraints for $[C_{\ell q}^{(1)}]_{3333}$ and $[C_{\ell q}^{(3)}]_{3333}$ evaluated on the
    matched Wilson coefficients. The last column contains the plots in the plane $(s[C_{\ell q}^{(1)}]_{3333}/\Lambda^2, s[C_{\ell q}^{(3)}]_{3333}/\Lambda^2)$, where the red line denotes where each model lies: the scalar and vector leptoquarks are correctly classified by
    their spin, whereas the $t$-channel vectors evade the classification.}
    \label{tab:models}
\end{table*}

In this appendix, we provide further technical details regarding the spinning sum rules discussed in Section~\ref{sec:bounds}. Following Ref.~\cite{Remmen:2020uze}, we consider the amplitude $\mathcal{A}_{\alpha\beta}(s,t)$, where $s$ and $t$ are the usual Mandelstam variables associated with an elastic scattering process of
two fermions whose quantum numbers (e.g., flavor and helicity) are collectively denoted by $\alpha$ and $\beta$.
Expanding the UV amplitude in partial waves of fixed total angular momentum, $a_{\alpha\beta}^{J}(s)$, and 
assuming vanishing boundary terms at infinity, one finds \cite{Remmen:2020uze}
\begin{multline}\label{eq: sum rules master equation}
    \lim_{s\rightarrow 0}\partial_s\mathcal{A}_{\overline \alpha\beta}(s,0) = -\lim_{s\rightarrow 0}\partial_s\mathcal{A}_{ \alpha\beta}(s,0) \\=8\int_{s_0}^\infty \dfrac{\dd s}{s^2}\left(\Im a_{\overline{\alpha}\beta}^{0}(s)-3\Im a_{\alpha\beta}^{1}(s)\right)\,,
\end{multline}
where $s_0$ is the characteristic mass-squared scale of the UV completion.\footnote{Note that our partial-wave coefficients are normalized differently from those of Ref.~\cite{Remmen:2020uze}.}
Dominance of spin-0 (spin-1) states in the UV theory implies that the left-hand side of Eq.~\eqref{eq: sum rules master equation} is positive (negative) definite for any choice of $\alpha$ and $\beta$.
This property can be exploited to derive sum rules among the flavor-dependent entries of the Wilson coefficients of $\psi^4$ SMEFT operators, or between different $\psi^4$ operators with the same fermionic content. For completeness, we report in the following the sum-rule bounds associated with SMEFT operators in the Warsaw basis, since in Ref.~\cite{Remmen:2020uze} different conventions are used.
In particular, the combinations of Wilson coefficients reported in Table \ref{tab:sum_rules}, when contracted with $\alpha_p \alpha_r^* \beta^*_s \beta_t$, are positive (negative) if the completion is dominated by scalars (vectors).
As a final remark, we stress that these bounds are subject to some limitations that may lead to their violation, as already discussed in the main text.

\section{Matching to UV completions\label{app:UV}}

To illustrate both the diagnostic power and the limitations of the spinning sum
rules discussed in the main text and App.~\ref{app: sum rules}, we match a
representative set of single-field, tree-level UV completions \cite{deBlas:2017xtg}
onto the semileptonic operators $[O_{\ell q}^{(1)}]_{3333}$ and
$[O_{\ell q}^{(3)}]_{3333}$, whose unitarity and sum-rule constraints are shown in
Fig.~\ref{fig:3plots} (right). The results are collected in Table~\ref{tab:models}.

The two scalar leptoquarks ($\omega_1$, $\zeta$) and the two vector leptoquarks
($\mathcal{U}_2$, $\mathcal{X}$) generate these operators through $s$-channel
exchange in the elastic $\ell q \to \ell q$ amplitude, and therefore enter the
dispersion relation of Eq.~\eqref{eq: sum rules master equation} with their physical
spin. Explicitly, the relevant combinations of Table~\ref{tab:sum_rules},
$[C_{\ell q}^{(1)}]_{3333} \pm [C_{\ell q}^{(3)}]_{3333}$, are both non-negative for
$\omega_1$ and $\zeta$, and both non-positive for $\mathcal{U}_2$ and $\mathcal{X}$,
so that the scalar (vector) leptoquarks are correctly placed in the scalar-
(vector-) dominated region of Fig.~\ref{fig:3plots} (right).

By contrast, the color-singlet vectors $\mathcal{B} \sim (1,1)_0$ and
$\mathcal{W} \sim (1,3)_0$ couple separately to the lepton and quark currents and are exchanged in the $t$-channel. The boundary terms then no longer vanish, modifying Eq.~\eqref{eq: sum rules master equation}, and the sum-rule
classification fails: $\mathcal{W}$ yields combinations of opposite sign and thus
lies outside both the scalar- and vector-dominated regions, while $\mathcal{B}$ can
fall in the region associated with the opposite (scalar) spin, depending on the sign
of $[g_{\mathcal B}^\ell]_{33}[g_{\mathcal B}^q]_{33}$. These cases provide explicit
examples of the statement made in the main text that vector $t$-channel completions
can violate the spinning sum-rule constraints, whereas all tree-level scalar
completions satisfy them.

\bibliographystyle{JHEP}
\bibliography{letter_bib}
 
\end{document}